\documentclass[]{spie}  

\usepackage{amsmath,amsfonts,amssymb}
\usepackage{graphicx}
\usepackage[colorlinks=true, allcolors=blue]{hyperref}

\title{CUBES and its software ecosystem: instrument simulation, control, and data processing}

\author[a]{Giorgio Calderone}
\author[a]{Roberto Cirami}
\author[a,b]{Guido Cupani}
\author[a]{Paolo Di Marcantonio}
\author[a]{Mariagrazia Franchini}
\author[c]{Matteo Genoni}
\author[d]{Mikołaj Kałuszyński}
\author[c,e]{Marco Landoni}
\author[f]{Florian Rothmaier}
\author[c]{Andrea Scaudo}
\author[d]{Rodolfo Smiljanic}
\author[f]{Ingo Stilz}
\author[f]{Julian Stürmer}
\author[g]{Orlando Verducci}
\affil[a]{INAF - Osservatorio Astronomico di Trieste, via G. B. Tiepolo 11, 34131 Trieste, Italy}
\affil[b]{IFPU - Institute for Fundamental Physics of the Universe, via Beirut 2, I-34151 Trieste, Italy}
\affil[c]{INAF - Osservatorio Astronomico di Brera, via E. Bianchi 46 Merate (LC) 23807-Italy}
\affil[d]{Nicolaus Copernicus Astronomical Center, Polish Academy of Sciences, ul. Bartycka 18, 00-716, Warsaw, Poland}
\affil[e]{INAF - Osservatorio Astronomico di Cagliari, via della Scienza 5, Selargius (CA) - Italy}
\affil[f]{LSW - Landessternwarte, Zentrum für Astronomie der Universität Heidelberg, Königstuhl 12, 69117, Heidelberg, Germany}
\affil[g]{LNA - Laboratorio Nacional de Astrofisica, Rua Estados Unidos 154, Itajuba-MG, Brazil}

\authorinfo{Further author information: (send correspondence to Giorgio Calderone)\\Giorgio Calderone: E-mail: giorgio.calderone@inaf.it}

\begin{document} 
\maketitle

\begin{abstract}
CUBES (Cassegrain U-Band Efficient Spectrograph) is the recently approved high-efficiency VLT spectrograph aimed to observe the sky in the UV ground-based region (305-400 nm) with a high-resolution mode ($\sim$~20K) and a low-resolution mode ($\sim$~5K). In this paper we will briefly describe the requirements and the design of the several software packages involved in the project, namely the instrument control software, the exposure time calculator, the end-to-end simulator, and the data reduction software suite.  We will discuss how the above mentioned blocks cooperate to build up a {\it ``software ecosystem''} for the CUBES instrument, and to support the users from the proposal preparation to the science-grade data products.
\end{abstract}


\section{INTRODUCTION}
\label{sec:intro}  

CUBES (Cassegrain U-Band Efficient Spectrograph) is the recently approved VLT spectrograph aimed to observe the sky in the UV ground-based region (305-400 nm, goal is 300-420 nm) with a high-resolution mode ($\sim$~20K) and a low-resolution mode ($\sim$~7K).  The most important feature of CUBES is its high efficiency in the UV, up to 40\%, which is significantly higher than other instruments observing in the same wavelength range (e.g. VLT/UVES).  Another important aspect is related to the software, since CUBES will be one of the first instruments adopting the recently released ELT-SW framework and standards, which will be used to develop and control all the future ESO/ELT instruments. Still, the CUBES instrument will ``live'' in a VLT environment since it will be installed at the Cassegrain focus of one of the VLT telescopes at the Paranal observatory (which of the four UT will be used is yet to be decided), and a specific software component, dubbed VLT / ELT gateway (\S\ref{sec:gateway}), is foreseen to allow communication between the CUBES software and the VLT environment. Finally, an important synergy is established with the VLT/UVES instrument, since CUBES will be able to operate UVES in order to perform simultaneous observations (using the red arm of UVES, operating in the 420-1100 nm wavelength range).

\medskip

\begin{figure}[ht]
  \begin{center}
    \begin{tabular}{c} 
      \includegraphics[width=\textwidth]{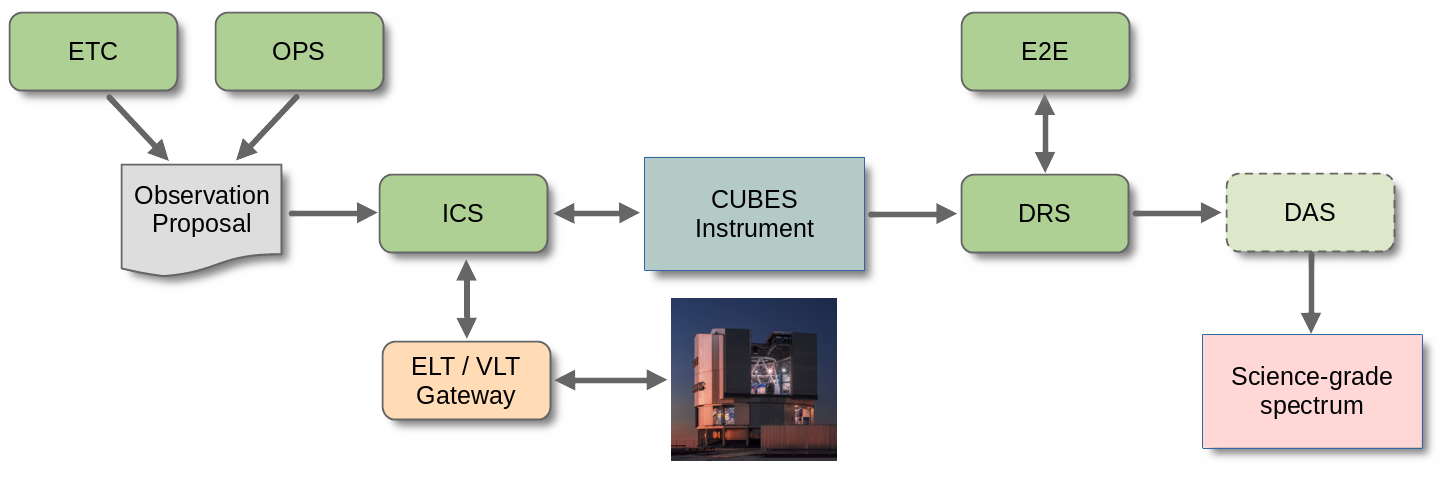}
    \end{tabular}
  \end{center}
  \caption[swecosystem]
          { \label{fig:swecosystem} 
            The CUBES instrument (box in the middle) surrounded by its {\it ``software ecosystem''}.}
\end{figure}

A description of the instrument is outside the scope of this paper (see [\citenum{2018-Evans_SPIE}] and [\citenum{2022-Cristiani_SPIE}] and references therein for a comprehensive overview of the instrument and of the science cases). Here we will focus on the software packages being developed by the CUBES consortium for the instrument exploitation:
%
\begin{itemize}
\item The Instrument Control Software (ICS, \S\ref{sec:ics}): used to control all the instrument functions and detectors, as well as to coordinate the observations and save the results in a form suitable to be ingested in the archive;
\item The Exposure Time Calculator (ETC, \S\ref{sec:etc}): used to estimate the exposure time required to achieve a given Signal-to-Noise Ratio (SNR);
\item The Observation Preparation Software (OPS, \S\ref{sec:ops}): used to collect all required information to prepare the observations; 
\item The Data Reduction Software (DRS \S\ref{sec:drs}): used to remove instrument artifacts from the science exposures, and extract a 1D spectrum;
\item The End-to-end Simulator (E2E, \S\ref{sec:e2e}): used to simulate the instrument behaviour as a whole and allowing early development of the DRS (using simulated images).
\end{itemize}
The above components cooperate to create a {\it ``software ecosystem''} around the instrument (see Fig.~\ref{fig:swecosystem}) aimed to support the instrument design before the actual hardware is available (E2E) as well as to support the users during operations, with applications ranging  from the initial proposal preparation (ETC and OPS), to the instrument control and operations (ICS, VLT/ELT Gateway) and the final data reduction and analysis (DRS) to obtain the science-grade data products.
A brief description of such components will be given in the following sections.

\section{Instrument Control Software (ICS)}
\label{sec:ics}
\begin{figure}[ht]
  \begin{center}
    \begin{tabular}{c} 
      \includegraphics[width=.8\textwidth]{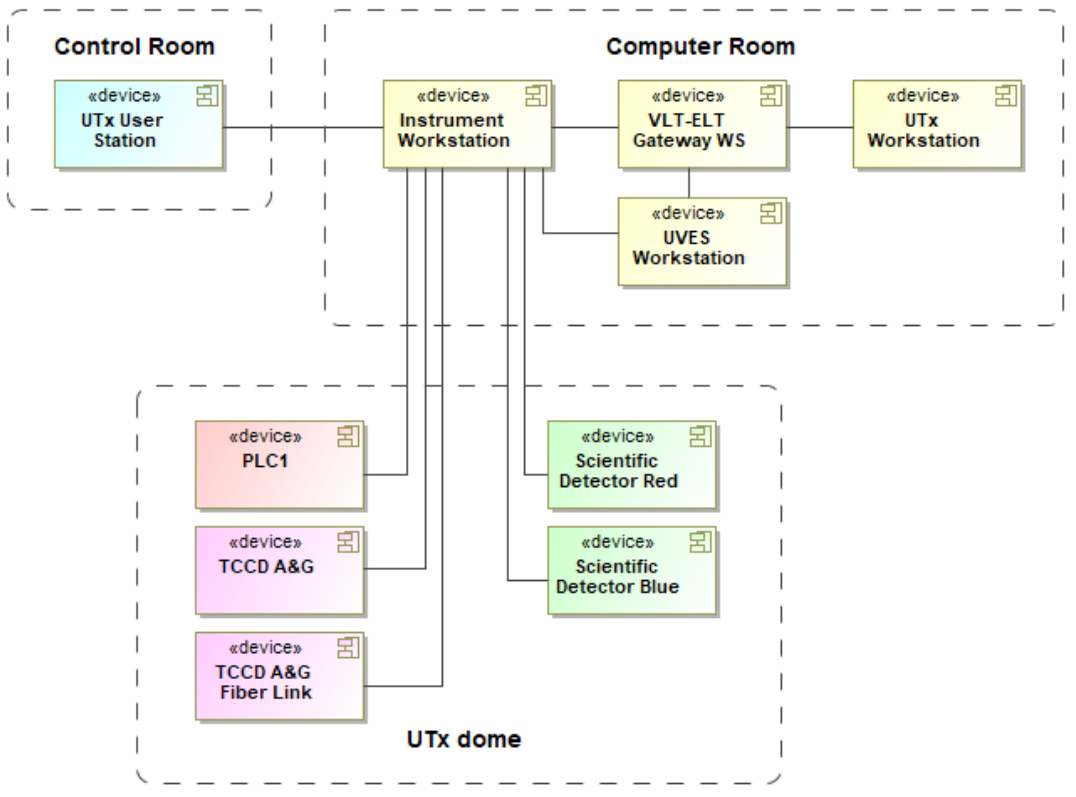}
    \end{tabular}
  \end{center}
  \caption[controlhw]
          { \label{fig:controlhw} 
            The CUBES control nodes and their connections.}
\end{figure}

The Instrument Control Software (ICS) is responsible to control and operate the whole CUBES instrument, as well as to collect the data and deliver the final FITS file to the ESO archive.  It can be conceptually divided into ``Low level'' (hardware control, \S\ref{sec:ics-low}), ``High level'' (data collection and preparation of the final FITS file \S\ref{sec:ics-high}) and ``Templates'' (scripts to perform instrument maintenance, science target acquisition and observations,  \S\ref{sec:ics-templates}). The ICS software operates on several devices, which are physically located in different rooms at the Paranal observatory (see Fig.~\ref{fig:controlhw}):
\begin{itemize}
\item The UTx  user station is the terminal used to operate the instrument, and is located in the same zone as the telescope terminal (``control room'').  Both the instrument operator and telescope operator collaborate to carry out the observations during the night.  In day time the same terminal can be used to run the calibration procedures.  No CUBES software is actually executed on the user station, it is just used as a terminal to access the instrument workstation;
\item The workstations to operate the instrument are located (among many other ones) in the ``computer room''.  More specifically, a dedicated instrument workstation will be employed to run all the CUBES ICS software (see sections below), and another one is used to run the VLT/ELT gateway (\S\ref{sec:gateway}).  The UVES workstation and the UTx workstations (both accessible through the gateway) are also located in the same room.  It is currently under discussion whether the gateway should have a dedicated workstation, or if it should be executed on a virtual machine hosted in the CUBES instrument workstation;
\item The PLCs (Programmable Logic Controllers) are used to execute the low level software for hardware control and monitoring, and are attached to the body of the instrument.  The technical cameras and the science detectors, located on the instrument itself, are also shown for completeness.
\end{itemize}

The most relevant requirement for the CUBES ICS is the one concerning the adoption of the ELT-SW framework and standards, which is currently under development at ESO and will be adopted in all the future instruments for the ELT telescope.  Such requirement affect all the ICS software and will be briefly discussed in the next sections. It is worth to recall that ESO ELT-SW is still under development, and several details may change in the future.

\subsection{Low level control software}
\label{sec:ics-low}
The low level control software is in charge to control all the instrument functions such as motors and actuators (20), calibration lamps (5), as well as to monitor the sensors (number is yet to be defined).  Hardware control is performed though a single standard FCS (Function Control System) device manager, provided by the ELT-SW, and a set of YAML configuration files for each of the devices to be controlled.  No ``special device'' (i.e. a device requiring dedicated extension of the device manager) is currently foreseen for CUBES. A dedicated configuration for each device is also foreseen on the PLCs.  Finally, the low level control software will also provide a device simulator, one for each actual device, which is able to emulate the PLC state machine and operational behavior even in the absence of a real hardware PLC.  Such simulators are very important to simulate the overall instrument behavior from the control software point of view, and are fundamental to develop and test the templates (\S\ref{sec:ics-templates}) well before the hardware becomes available.

\medskip

The CUBES instrument also features two technical cameras (one to measure deviations of the light path from the fiber feeding the UVES instrument, and a second one used to perform target acquisition and guiding), and two science detectors (one for each arm).  These devices will be controlled respectively by the ELT-SW Camera Control Framework (CCF) and the New General Detector Controller II (NGCII) frameworks.

\subsection{High level control software}
\label{sec:ics-high}

The high level control software comprises many packages (and the associated configurations) involved in the orchestration and data collection of the several CUBES subsystems, as dictated by the new ESO ELT-SW standard.  Worth mentioning are the the Observation Coordination Manager (OCM) which coordinates the data acquisition by collecting the images from the science detectors and triggers the creation of metadata from all the instrument subsystems, and the Data Product Manager (DPM) which prepare and write the final FITS file, ready to be transferred to the On-Line Archive Subsystem (OLAS). Another relevant high level component  is the sequencer, devoted to the execution of the Observation Blocks (i.e. collections of one or more templates, see \S\ref{sec:ics-templates}) queued in the observation schedule.  The sequencer operates by reading the JSON file describing a template execution, including the reference to the Python script to be executed as well as the specific parameter values, and orchestrate the dispatch of the proper messages to the CUBES subsystem (FCS device manager, OCM, CCF and NGCII).  The sequencer is also used to execute engineering scripts for maintenance of the instrument, or to start/stop the whole instrument software). Finally, the templates represent the most important part of the high level control software, and are therefore described in a separate section below.

\subsection{Templates}
\label{sec:ics-templates}


As mentioned in \S~\ref{sec:ics-high}, the execution of instrument actions is organized in the form of templates. In the ELT-SW, a template is a Python script aiming to operate the individual instrument devices in the correct sequence. According to their range of use, templates are subdivided into the following four groups:

\begin{itemize}
  \item acquisition templates: in charge of acquiring the field of observation; this requires communication with the Telescope Control Software in order to initiate the standard telescope procedures like slewing, tracking and guiding;
  \item observation templates: the centerpiece of the instrument operation with the purpose of taking science data;
  \item calibration templates: responsible for carrying out the calibration of the instrument, particularly of the science and technical CCDs;
  \item maintenance templates: preventive actions to ensure the mid- and long-term health of the instrument.
\end{itemize}
A template contains the series of steps belonging to a clearly defined task. Templates are included in the so-called ``Observation Block'' (OB), corresponding to an individual observation in a proposal. Executing an OB means running one or several templates according to the operations schedule. In that sense, templates may be considered the bricks of an OB.

From the software's perspective, CUBES will be operated in two different modes: CUBES (standalone) and CUBES+UVES, These modes are reflected in the template design. The mode CUBES+UVES will be preferably operated in the way that UVES is remotely controlled from CUBES. A typical sequence could consist of the following steps:

\begin{enumerate}
  \item From the OB available on the CUBES workstation, an OB for UVES is created and transferred to the UVES workstation;
  \item Relevant database values are synchronized between the two workstations via a dedicated gateway;
  \item The OB on the UVES workstation is remotely executed;
  \item An exposure with CUBES is taken;
  \item CUBES waits until the UVES exposure has been completed.
\end{enumerate}
The number of templates as currently foreseen for the four template groups is given in Table~\ref{tab:cubes_no_templates}.

\begin{table}[ht]
\caption{Number of templates foreseen for the individual groups.} 
\label{tab:cubes_no_templates}
\begin{center}       
\begin{tabular}{|l|l|} 
    \hline
    Template type & No. of templates \\
    \hline
    Acquisition & 3 \\
    Observation & 2 (CUBES and CUBES + UVES) \\
    Calibration & 7 \\
    Maintenance & 7 \\
\hline 
\end{tabular}
\end{center}
\end{table}

\subsection{The VLT/ELT gateway}
\label{sec:gateway}

CUBES will be one of the first instruments to adopt ELT-SW framework and standards, which will be used to develop and control all the future ESO/ELT instruments. But the CUBES instrument operates in a VLT environment (Paranal observatory), hence the communications between CUBES and its environment needs to be mediated by a dedicated software component, dubbed VLT / ELT gateway.

\medskip

The main functionality exposed by the gateway is the possibility to automatically translate the communication protocols from a VLT environment to an ELT one, and vice-versa.  This means that a message sent with the new ELT-SW sequencer could be received as if it was issued by an old VLT CCS application, and that a change in a VLT On-Line Database (OLDB) could be reflected in a ELT application database.  The above examples are exactly the use cases required for CUBES, i.e. the possibility to send commands to the TCS telescope system (for target pointing) and to UVES (to start the simultaneous exposure), as well as the possibility to read the status of TCS and UVES in their OLDBs.

\section{Exposure Time Calculator (ETC)}
\label{sec:etc}

The Exposure Time Calculator (ETC) is a tool to predict the performances of the CUBES spectrograph in different environmental conditions. 
It is a web-based tool through which the science community, in synergy with the E2E simulator (see \S~\ref{sec:e2e} and [\citenum{GEN22}]),  has been able to test a range of science cases for CUBES during the phase-A. 
The program supports two modes of functionality: at a given wavelength ($\lambda$) and for a given target magnitude in the $U$ or $V$ bands it computes (1) the SNR$_{\lambda}$ per pixel achievable with a given user input exposure time ($t_{exp}$) and (2) the exposure time required to obtain a given  user input SNR$_{\lambda}$ per pixel. Finally, it also allows to compute the limiting magnitude (in AB units)  achievable at a pre-selected wavelength (assuming a SNR$_{\lambda}$=3).

\begin{figure*}
  \includegraphics[width=0.5\textwidth,angle=0]{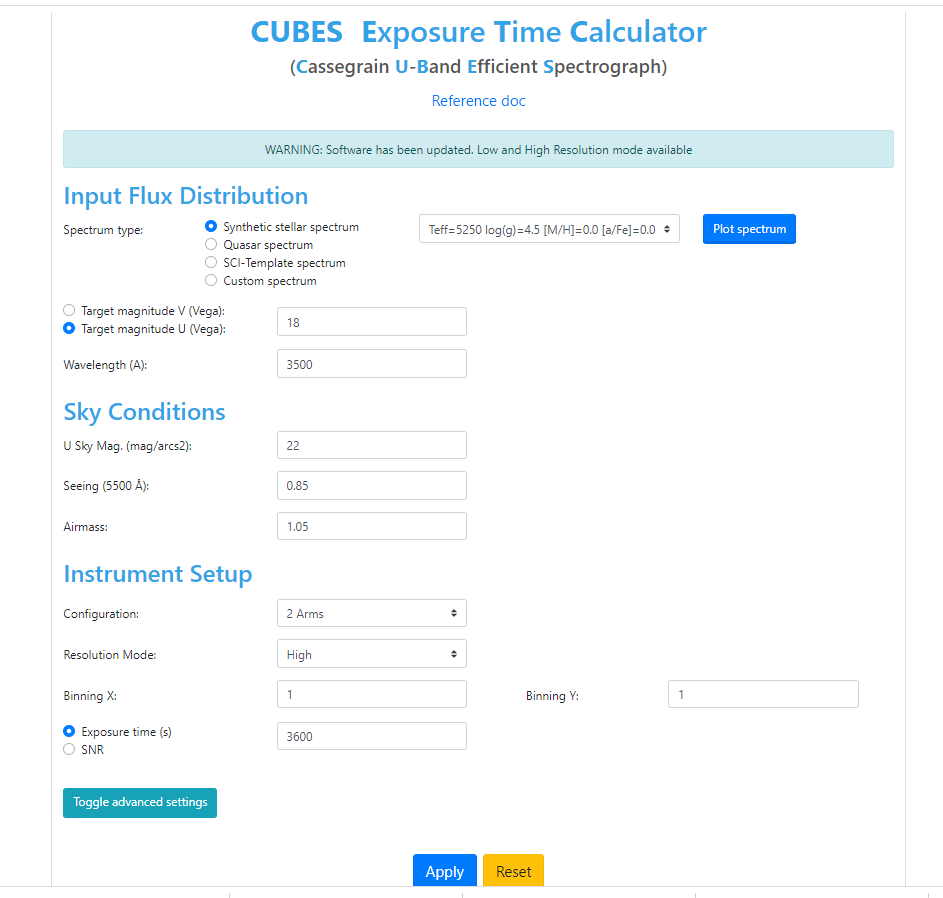}
  \includegraphics[width=0.56\textwidth,angle=0]{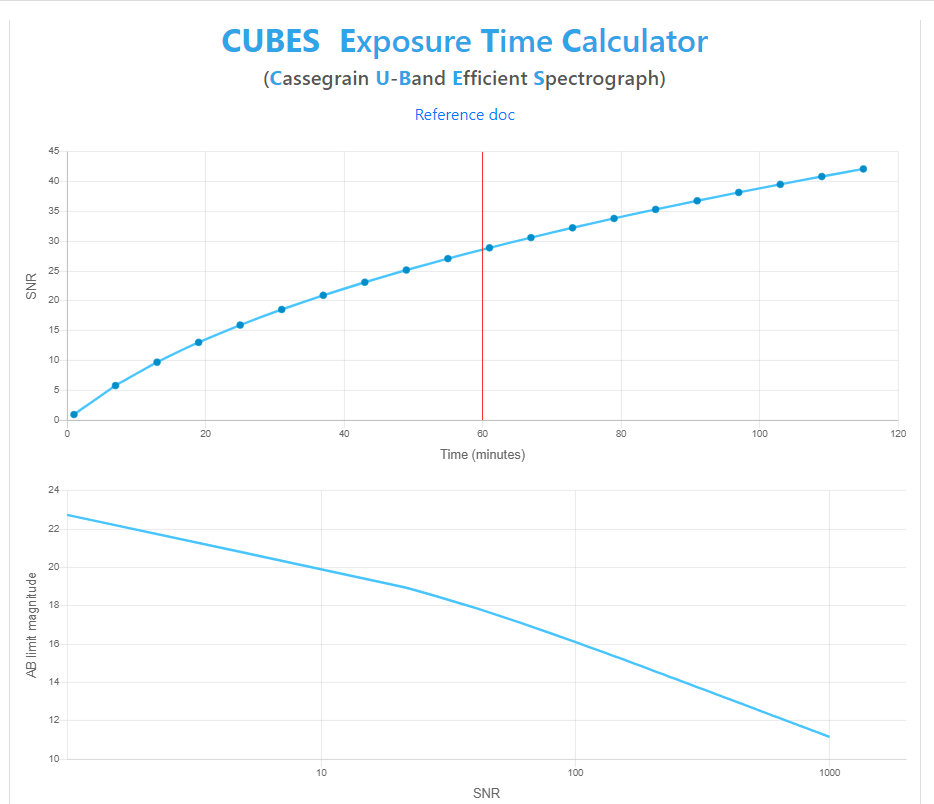}
\caption{Left panel: first web-page of the CUBES ETC, showing the inputs required to compute SNR$_\lambda$. Right panel: graphical outputs of the CUBES ETC version at the end of Phase~A. In the top panel is plotted the predicted SNR$_\lambda$ vs t$_{exp}$  at $\lambda$=350\,nm for different exposure times with the high resolution mode of CUBES; the input template is a synthetic spectrum (effective temperature $T_{\rm eff}$=5250\,K, gravity $\log g$=4.5, iron abundance [Fe/H]=0, and $\alpha$-elements abundance [$\alpha$/Fe]=0) available in the ETC database as template  and scaled to  $U$ = 18\,mag, with other parameters set to: $U_{\rm Sky}$ = 22\,mag\,arcsec$^{-2}$, airmass = 1.05, seeing = 0.85$''$,  dark current = 0.5e$^-$/pix/hr, RON = 2.5\,e$^-$ rms, and 1\,$\times$\,1 pixel binning; the vertical red line shows the SNR$_{\lambda}$ at the selected exposure time of 1\,h. In the bottom panel is plotted  SNR$_{\lambda}$ vs  AB magnitude limit diagram computed for $t_{exp}=1$\,h and the same conditions as in the top panel. }
\label{fig:ETC_input}       
\end{figure*}

The ETC is a Python 3.7.6 script that is called by a web
application written in Java and JavaScript and uses the
facilities of the Italian Center for Astronomical Archive
(IA2) operated by INAF\footnote{\url{http://archives.ia2.inaf.it/cubes/\#/etc}}. 
The main page of ETC (left panel of Fig.~\ref{fig:ETC_input}) is the observation parameters page and  presents the entry fields and widgets for the target information ($U$ or $V$ Vega magnitude, type of input spectrum), expected atmospheric
conditions (the $V_{sky}$ magnitude in mag$\cdot$arcsec$^{-2}$, seeing, airmass), instrument configuration (low or high resolution mode, CCD binning), and the  exposure time or signal-to-noise input parameter depending on the required output (SNR$_{\lambda}$ per pixel or  $t_{exp}$). The CUBES
setup (i.e. slit, image slicer, gratings, cameras and CCDs characteristics) is hidden to the user, but can be visualized via the ‘Toggle advanced setting’ at the bottom of the main page and the setup parameters can be modified within their acceptable ranges allowing us to explore different design solutions during the Phase A study (for more details, see [\citenum{GEN22}]). The result page   presents the computed results, i.e. two graphical outputs (SNR$_{\lambda}$ vs. $t_{exp}$ and the limiting magnitude vs. SNR$_{\lambda}$ as shown in the top and bottom plots, respectively in the right panel of   Fig.~\ref{fig:ETC_input}). In addition to these graphical outputs, a table including the main input data
(like object type, sky condition, slit losses, number
of counts from the object and the sky, wavelength bin
in \AA/pix, instrument efficiencies, detector binning,
detector noise counts, magnitude limit at SNR$_{\lambda}$ = 3,
and SNR/$t_{exp}$) is provided.

Current version of ETC computes SNR$_{\lambda}$ only for point source. We use  the following classical equation for computing the  CUBES SNR$_{\lambda}$:
\begin{equation}
{SNR_{\lambda}=\frac{PE^{obj}\cdot t_{exp}}{\sqrt{[PE^{obj}+PE^{sky}]\cdot t_{exp} +{N_{DC}}^2\cdot t_{exp}+{N_R}^2}} }
 \label{eq:snr}
\end{equation}

\noindent where  $PE^{obj}$ and $PE^{sky}$ are the numbers of photo-electrons per second and per wavelength bin (i.e. \AA~ per pixel) from the target and the sky, respectively; 
${N_{DC}}^2$ is the detector noise due to the dark current in photo-electrons per second and  integrated by summing the counts on the CCD area defined by the integer number of pixels in the dispersion direction ($nx$=1) corresponding to the wavelength bin and by the number of pixels in the spatial direction ($ny$) depending on the observational conditions\footnote{In the current version of ETC we compute $ny$ by assuming of integrating the signal of each spectrum over a spatial region of 1.2 times the seeing.};  ${N_R}^2$ is the detector read-out noise in photo-electrons   per $nx\cdot ny$ pixels;      $t_{exp}$ is the integration time in seconds.

Using an appropriate parameterization,  ETC models the effects of the  atmosphere,   telescope,   slit,  spectrograph,  camera and  detector and includes their effects  on the target’s photons  that are derived starting from a template spectrum (from the local database or a user-uploaded template). The ETC modelling approach is coherent to that developed by the E2E simulator (\S~\ref{sec:e2e}) as confirmed by the several and careful comparisons between the E2E and ETC for several reference input configurations. 
The reliable computation of
SNR$_{\lambda}$ in a variety of  observing conditions, has provided a valuable tool for the
astronomical community to explore the wide range of
science cases that CUBES aims to address (see [\citenum{EVA22}]).

\section{Observation Preparation Software (OPS)}
\label{sec:ops}

The Observation Preparation Software (OPS) aims to assist the user in designing the best observing strategy to achieve a specific science goal. It guides the user through the allowed settings for the instrument (e.g. choice of target and guide stars, instrument setup, choice of auxiliary calibrations, etc.). We are currently investigating whether further CUBES-specific (so called ``Level 3 OPS'') functionalities are required to, e.g., assist the user in planning a monitoring campaign, or to implement a real-time scheduling strategy.

\section{Data Reduction Software (DRS)}
\label{sec:drs}

The Data Reduction Software (DRS) is responsible for extracting science-grade spectra from the raw science and calibration frames produced by the instrument. Its main capabilities are: removing instrumental signature; calibrate 2D spectra in physical units; remove sky and instrumental backgrounds as well as cosmic ray traces; extract a 1D spectrum (also combining different exposures); and propagate the uncertainties. The requirements on the DRS performances are: 5\% precision and 20\% accuracy in flux calibration; 1.5e-3 nm accuracy in wavelength calibration; execution time $<$ 8 hours (goal: $<$3 hours) to reduce an observing night; $<$5 minutes to reduce an OB for quality control.

The DRS is based on the ESO Common Pipeline Library (CPL) and the High-level Data Reduction Library (HDRL) frameworks. The DRS package will include:
\begin{itemize}
\item a pipeline of reduction recipes (i.e. individual procedures, suitable to be run in cascade);
\item a library of atomic reduction routines used by the recipes, to complement the CPL and the HDRL (if needed), provided with unit tests;
\item a set of configuration files specifying the observational setup of the instrument and the organization, classification, and association (OCA) rules for the input and output data frames;
\item a set of reference files specifying the spectral format, the position of lines for wavelength calibration, etc.
\item one or more Python workflows implementing the cascade (with specific configurations for science and quality control reduction), equipped with visualization tools to assess the status of the procedure;
\item one or more sets of demo data to perform test reductions.
\end{itemize}

\begin{figure}[ht]
  \begin{center}
    \begin{tabular}{c} 
      \includegraphics[width=0.6\textwidth]{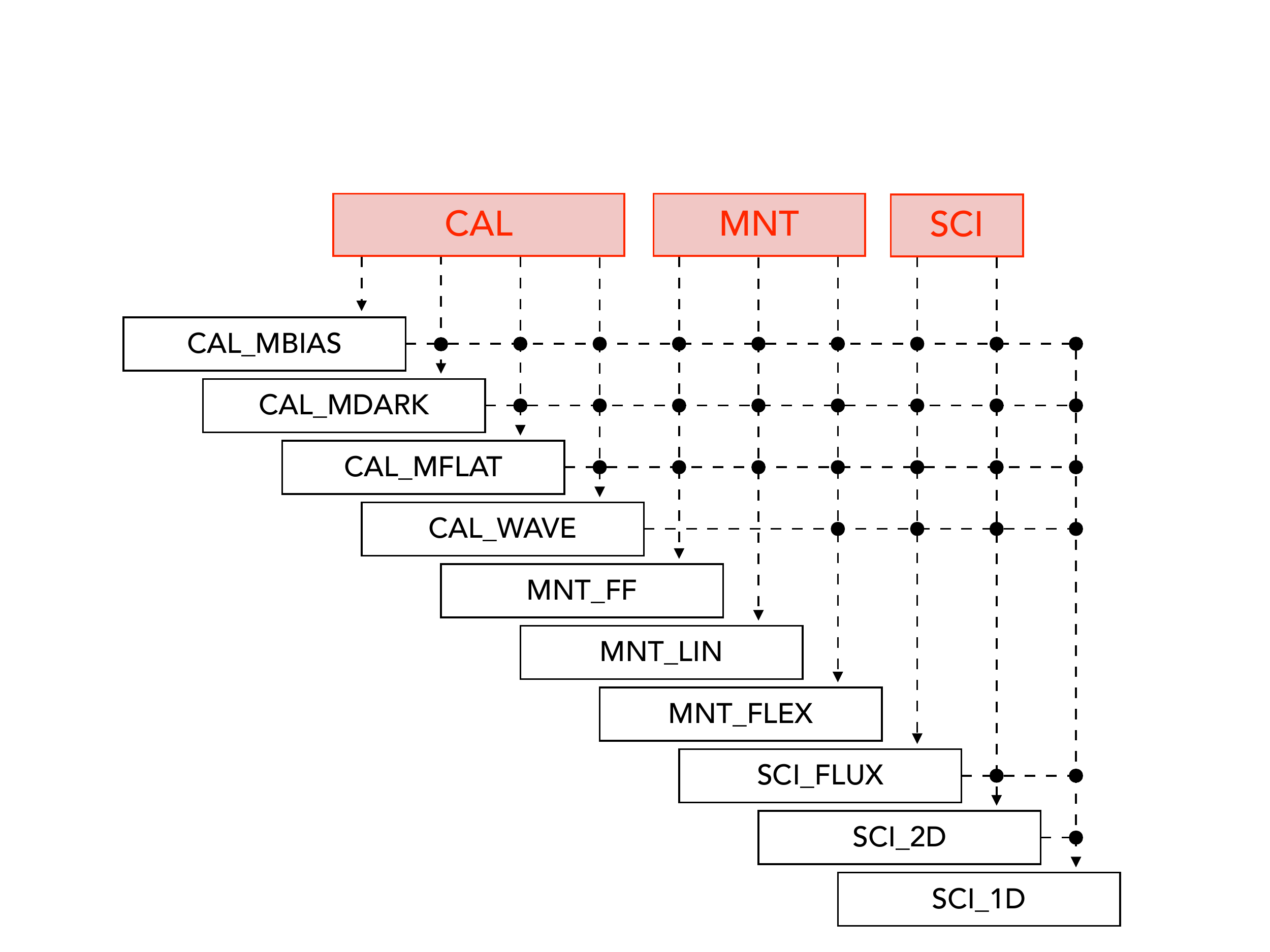}
    \end{tabular}
  \end{center}
  \caption[drs]
          { \label{fig:drs} 
            Pipeline of the CUBES Data Reduction Software, with association between recipe products.}
\end{figure}

A schematic representation of the DRS pipeline is given in Fig.~\ref{fig:drs}. Recipes are shown as white boxes and grouped into recipes for routine calibration (CAL, including master bias, master dark, and master flat creation, and wavelength calibration), maintenance (MNT, including detector flat-fielding, linearity tests, and flexure compensation) and science reduction (SCI, including flux calibration, 2D spectrum reduction and 1D spectrum extraction). The main input of recipes are the calibration and science frames acquired by the instrument detectors; the dotted lines show how products from previous recipes are associated to these raw frames to serve as input of subsequent recipes. Fig.~\ref{fig:drs} also highlights one key feature of the DRS, i.e.~the fact that the reduction of 2D spectra  and the extraction of 1D spectra are fully decoupled from each other. A raw spectrum is first reduced in the 2D space of the detector pixels by recipe SCI\_2D (performing bias and dark subtraction, flat-fielding, wavelength and flux calibration) and then extracted and rebinned into a custom wavelength grid as a 1D spectrum by recipe SCI\_1D (adopting a drizzling-like algorithm originally developed for the ESPRESSO Data Analysis Software; see [\citenum{Cupani16}]). This approach provides a natural way to selectively extract individual slices from science frames and to coadd several exposures into a single 1D spectrum with a single rebinning procedure.

The DRS package will be able to operate both offline (upon download on the users' machines) and online (in Paranal, triggered by the injection of data in the ESO  On-Line Archive Subsystem). In the CUBES+UVES mode, it will operate alongside the existing UVES data reduction pipeline.

\subsection{Data Analysis Software (DAS)}
\label{sec:das}

The Data Analysis Software (DAS) is a collection of science-focused procedures which operate in close interaction with the DRS and which are aimed to extract scientific information from the reduced spectra, minimizing the work load of the science user. It is currently under investigation whether such tools are beneficial for CUBES to, e.g., perform the co-addition of spectra or to extract specific information from a CUBES monitoring program.

\section{End-to-end Simulator (E2E)}
\label{sec:e2e}

The End-to-End Instrument simulator (E2E) aims to simulate CUBES behaviour as a whole. Its inputs are the observation conditions, the instrument setup and the target specifications, and its output is a raw data image which simulates an observation carried out at the aforementioned conditions. The main purpose of the E2E is to evaluate the performances of the instrument design, as well as to feed the DRS with realistic data in order to aid its development and verify the requirement compliance.
It has a modular architecture where different modules (each with specific tasks), units and interfaces, mimicking the functional architecture of the CUBES instrument (for more details see [\citenum{GEN22}]).
Three versions of the CUBES E2E have been developed for different user applications, featuring different capabilities (or functionalities) that were used during the Phase~A design study. 
These are:
\begin{itemize}
    \item \textit{Basic Version:} aimed at science users, to assess the performance of the instrument with respect to the science objectives, depending on the adopted instrument design.
    \item \textit{Parametric Version:} provides fast parametric simulations of the different possible design solutions and configurations, as an aid for the design development. 
    \item \textit{Full Version:} produces high-level simulated frames based on a detailed physical model of the instrument (including optical effects as the contribution of blurring of the point spread function (PSF), distortions and detector diffusion effects), to be used for the development and testing of the data reduction pipeline.
\end{itemize}

Given the different purposes and functionalities, the first version is available on-line for the scientific community, while the other two are used by the technical team.

The \textit{Basic Version} of the simulator is implemented as a Python 3 library and accessed through a Jupyter notebook [\citenum{jupyter}]. 
The notebook includes both the basic code to perform the simulation and instructions on how to execute it. 
Users can freely configure the parameters of the simulation and run it. 
The library and the notebook are available as a Git repository through GitHub\footnote{\url{https://github.com/gcupani/cubes}.} for download on local machines. It is also hosted on Binder [\citenum{binder}], an online tool to execute Jupyter notebooks on a remote server interactively\footnote{\url{https://bit.ly/cubes_e2e} (Note: the Binder version is limited in memory usage).}.
The \textit{Full Version, which} can be run both in MATLAB and Python (3.9), uses libraries for specific functionalities, as well as an ad-hoc software wrapper we developed to interface with other software, such as the commercial optical ray tracing Zemax-OpticStudio.
The \textit{Parametric Version}, currently developed only in MATLAB, shares some modules and parts of code with the \textit{Full Version} (e.g. the Science Object Module), while it employs an ad-hoc parametric description of the spectrograph to run parametric simulations.

For a detailed description of the different versions, their functionalities and how they were successfully used in Phase-A to aid instrument design choices (trade-off analyses of different potential designs for the HR mode, and to also assess the performance of a LR mode for the instrument), as well as  driving initial development of pipeline recipes for the DRS, see again [\citenum{GEN22}].

\begin{figure*}
\centering
  \includegraphics[width=0.85\textwidth]{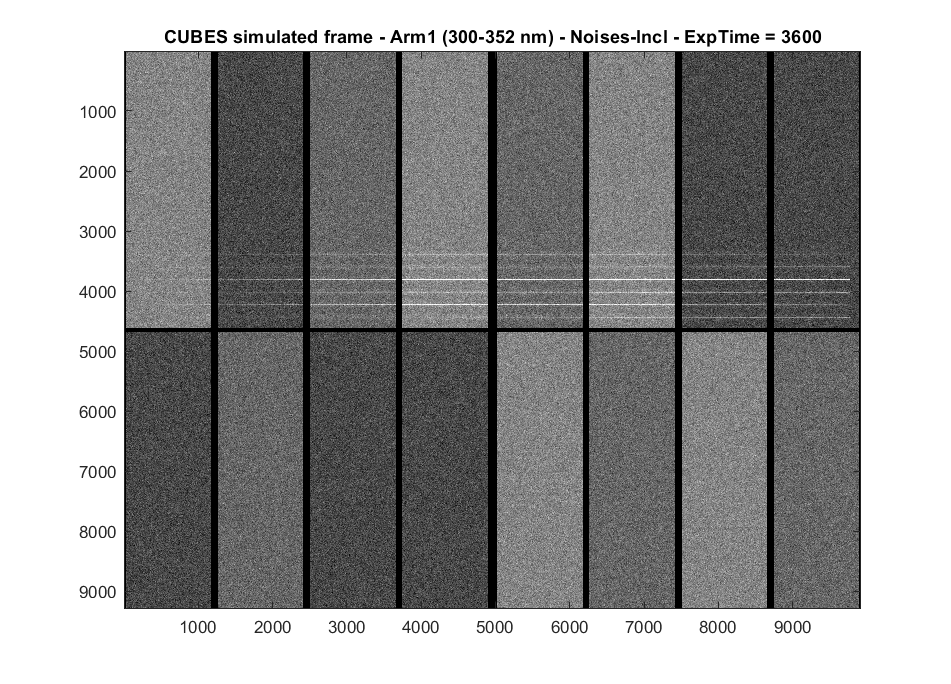}
\caption{Example simulated raw frame for Arm 1. The six slice traces are projected onto the upper half of the detector to allow the lower half to be read out separately to record the active flexure compensation (AFC) system (if analysis in the next phases shows it is required).}
\label{fig:E2E_Tech_07_Full_Frame_Arm1} 
\end{figure*}

\begin{figure*}
\centering
  \includegraphics[width=0.75\textwidth]{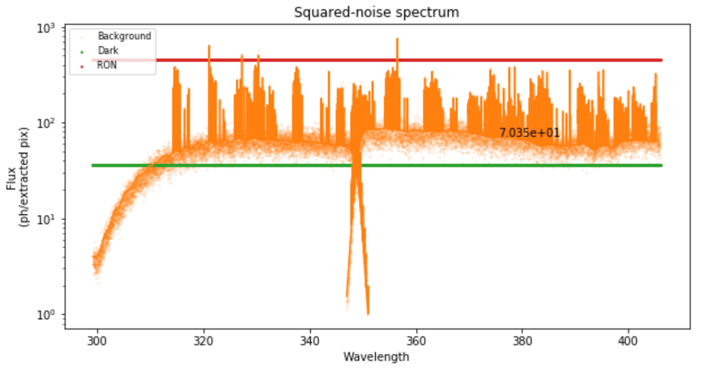}
\caption{Example of noises spectrum extracted per pixel bin, over the six slices, for a reference observing condition (see text for details).}
\label{fig:E2E_Sci_Noises} 
\end{figure*}

An example of a synthetic raw frame produced by the \textit{Full Version} is shown in Fig.~\ref{fig:E2E_Tech_07_Full_Frame_Arm1}. 
The traces of the six slices can be seen, in which it is also possible to see the decreasing flux of the object towards the blue end of the spectral range (left side of the figure). 
In the baseline design the traces are projected onto the upper half of the detector because the lower half is used to record the AFC fibre (if required).
In the reference observing condition for the HR mode, i.e. new moon, airmass 1.16, precipitable water vapour 30 mm and seeing = 0.87$''$ in 1 hour exposure, the estimated detector noise (dark current = 0.5e$^-$/pix/hr, RON = 2.5\,e$^-$ rms) dominates over the sky background and the sky flux in the pixels cannot be easily distinguished; this is shown in Fig. \ref{fig:E2E_Sci_Noises}.

\acknowledgments

R.S. acknowledges support by the Polish National Science Centre through project 2018/31/B/ST9/01469.

\bibliography{main} 

\begin{thebibliography}{1}

\bibitem{2018-Evans_SPIE}
{Evans}, C.~J., {Barbuy}, B., {Castilho}, B., {Smiljanic}, R., {Melendez}, J.,
  {Japelj}, J., {Cristiani}, S., {Snodgrass}, C., {Bonifacio}, P., {Puech}, M.,
  and {Quirrenbach}, A., ``{Revisiting the science case for near-UV
  spectroscopy with the VLT},'' in [{\em Ground-based and Airborne
  Instrumentation for Astronomy VII}{\nolinebreak\hspace{0.1em}]},  {Evans},
  C.~J., {Simard}, L., and {Takami}, H., eds., {\em Society of Photo-Optical
  Instrumentation Engineers (SPIE) Conference Series} {\bf 10702},  107022E
  (July 2018).

\bibitem{2022-Cristiani_SPIE}
{Cristiani}, S. et~al., ``Cubes, the cassegrain u-band efficient
  spectrograph,'' {\em SPIE Conference series} {\bf 12189} (2022).

\bibitem{GEN22}
{Genoni}, M., {Landoni}, M., {Cupani}, G., {Franchini}, M., {Cirami}, R.,
  {Zanutta}, A., {Evans}, C., {Di Marcantonio}, P., {Cristiani}, S., {Trost},
  A., and {Zorba}, S., ``{The CUBES instrument model and simulation tools},''
  {\em Experimental Astronomy}  (Mar. 2022).

\bibitem{EVA22}
{Evans}, C.~J. and { et al.}, ``{The CUBES Science Case},'' {\em Experimental
  Astronomy (in press)}  (2022).

\bibitem{Cupani16}
{Cupani}, G., {D'Odorico}, V., {Cristiani}, S., {Gonz{\'a}lez-Hern{\'a}ndez},
  J.~I., {Lovis}, C., {Sousa}, S., {Calderone}, G., {Cirami}, R., {Di
  Marcantonio}, P., and {M{\'e}gevand}, D., ``{Integrated data analysis in the
  age of precision spectroscopy: the ESPRESSO case},'' in [{\em Software and
  Cyberinfrastructure for Astronomy IV}{\nolinebreak\hspace{0.1em}]},
  {Chiozzi}, G. and {Guzman}, J.~C., eds., {\em Society of Photo-Optical
  Instrumentation Engineers (SPIE) Conference Series} {\bf 9913},  99131T (July
  2016).

\bibitem{jupyter}
Kluyver, T., Ragan-Kelley, B., P{\'e}rez, F., Granger, B., Bussonnier, M.,
  Frederic, J., Kelley, K., Hamrick, J., Grout, J., Corlay, S., Ivanov, P.,
  Avila, D., Abdalla, S., Willing, C., and development team, J., ``Jupyter
  notebooks ? a publishing format for reproducible computational workflows,''
  in [{\em Positioning and Power in Academic Publishing: Players, Agents and
  Agendas}{\nolinebreak\hspace{0.1em}]},  Loizides, F. and Scmidt, B., eds.,
  87--90, IOS Press (2016).

\bibitem{binder}
{P}roject {J}upyter, {M}atthias {B}ussonnier, {J}essica {F}orde, {J}eremy
  {F}reeman, {B}rian {G}ranger, {T}im {H}ead, {C}hris {H}oldgraf, {K}yle
  {K}elley, {G}ladys {N}alvarte, {A}ndrew {O}sheroff, {P}acer, M., {Y}uvi
  {P}anda, {F}ernando {P}erez, {B}enjamin~{R}agan {K}elley, and {C}arol
  {W}illing, ``{B}inder 2.0 - {R}eproducible, interactive, sharable
  environments for science at scale,'' in [{\em {P}roceedings of the 17th
  {P}ython in {S}cience {C}onference}{\nolinebreak\hspace{0.1em}]},  {F}atih
  {A}kici, {D}avid {L}ippa, {D}illon {N}iederhut, and {P}acer, M., eds.,  113
  -- 120 (2018).

\end{thebibliography}
\bibliographystyle{spiebib} 

\end{document}